\documentstyle[12pt]{article} 
\font\tenrm=cmr10 

\newcommand{\bref}[1]{(\ref{#1})} 
\newcommand{\ct}[1]{\cite{#1}} 
\def\plusstate{| \left. + \right\rangle} 
\def\state{| \left. s \right\rangle} 
\def\stateprime{| \left. { s' } \right\rangle} 
\def\kstate{| \left. k \right\rangle} 

\newcommand{\be}{\begin{equation}} 
\newcommand{\ee}{\end{equation}} 

\begin{document}  
\titlepage  
  
\begin{flushright}  
{CCNY-HEP-97/4\\}   
{CU-TP-829\\} 
{hep-th/9705167\\}  
{May 1997\\}  
\end{flushright}  
\vglue 1cm

\begin{center}   
{ 
{\Large \bf Solutions of Extended Supersymmetric \\  
Matrix Models for Arbitrary Gauge Groups     
\\}  
\vglue 1.0cm  
{Stuart Samuel\\}   
\bigskip

\medskip 

{\it Department of Physics$^{*}$\\}
{\it City College of New York\\}
{\it New York, NY 10031, USA\\} 

{\it and \\} 

{\it Department of Physics\\}
{\it Columbia University\\}
{\it New York, NY 10027, USA\\} 
\vglue 0.8cm  
} 
 
\vglue 0.3cm  
  
{\bf Abstract} 
\end{center}  
{\rightskip=3pc\leftskip=3pc 
\quad Energy eigenstates for 
$N=2$ supersymmetric gauged quantum mechanics 
are found 
for the gauges groups $SU(n)$ and $U(n)$. 
The analysis 
is aided by the existence 
of an infinite number of conserved operators. 
The spectum is continuous. 
Perturbative eigenstates 
for $N>2$ are also presented, 
a case which is relevant 
for the conjectured description 
of M theory in the infinite momentum frame.  

} 

\vfill
 
\textwidth 6.5truein
\hrule width 5.cm
\vskip 0.3truecm 
{\tenrm{
\noindent 
$^*$Permanent address.\\   
\hspace*{0.2cm}E-mail address: samuel@scisun.sci.ccny.cuny.edu\\}}
 
\newpage  
  
\baselineskip=20pt  
 
{\bf\large\noindent I.\ Introduction}  

A view has emerged that different superstring theories 
are various limits of one unifying theory.   
That theory is M theory, 
an eleven-dimensional system whose low-energy limit 
is $D=11$ supergravity and whose 
compactifications to ten dimensions 
on a circle and an interval 
yield type IIA and $E_8 \times E_8$ heterotic 
superstrings.\ct{dhis87}-\ct{schwartz96}  
Dualities  \ct{schwartz96}-\ct{duff96b}, 
as well as new string degrees of freedom 
such as D branes \ct{polchinski95,polchinski96} 
and supermembranes \ct{hlp86,townsend92,duff96c}, 
have played a role in obtaining 
this unification. 
Supersymmetric matrix quantum mechanics  
has been useful in gaining insight into 
D-branes, supermembranes and M theory.\ct{whn88}-\ct{bfss96}  
Although a precise {\it covariant} formulation of M theory 
is still lacking, 
it has been conjectured 
that the $n \to \infty$ limit 
of an $SU(n)$-matrix quantum mechanics system 
with $N=16$ supersymmetry 
describes M theory 
in the infinite momentum frame.\ct{bfss96} 
Hence, 
any progress in understanding such systems 
is of interest. 

Ground states of supersymmetry quantum mechanics 
are often of the from $\exp (\pm W)$, 
where $W$ is the superpotential.\ct{witten81,ch85}. 
Excited energy states are not known 
with one exception: 
All such states of the $N=2$ supersymmetric $SU(2)$ gauge theory 
have been found.\ct{ch85} 
In this letter, 
we obtain energy eigenstates for the $U(n)$ and $SU(n)$ systems 
for general $n$. 
Given the interest in the $SU(\infty)$ case, 
out results  should be of use in future research. 
The $N=16$ case, 
which is relevant for M theory 
is not solved. 
However, 
perturbative $N=16$ energy eigenstates, 
for which the coupling constant is set to zero, 
are found; 
this is non-trivial due to the requirement 
of satisfying the Gauss law constraints. 
Our results are a first step toward a perturbative analysis 
of the M theory proposal of 
ref.\ct{bfss96}. 

\medskip 
{\bf\large\noindent II.\ Particular Solutions for the $N=2$ Case}

The $N=2$ quantum-mechanic gauge theory 
involves a real gauge potential $A_B$, 
a real scalar $\phi_B$, 
and a complex fermion $\psi_B$, 
all in the adjoint representation.  
Here, $B$, which is a gauge index, 
runs over the number $n_G$ 
of generators of the Lie group $G$, 
e.g.\ $n_G = n^2-1$ for $SU(n)$ 
and $n_G = n^2$ for $U(n)$. 
The lagrangian ${\cal L}$ 
is 
\be  
  {\cal L} = 
  {1 \over 2}\left( {{\cal D}_t\phi } \right)_A 
  \left( {{\cal D}_t\phi } \right)_A + 
  i\bar\psi_A \left( {{\cal D}_t\psi } \right)_A - 
  ig f_{ABC} \bar\psi_A \phi_B \psi_C 
\quad , 
\label{eq1} 
\ee 
where the covariant derivative ${\cal D}_t$ 
on any field $\varphi$ is 
$  
  \left( {{\cal D}_t\varphi } \right)_A \equiv 
   \partial_t \varphi_A - gf_{ABC} A_B \varphi_C 
$ 
and where $g$ is the gauge coupling. 
Here and elsewhere, 
the presence of a repeated index 
indicates summation. 
The $f_{ABC}$ are structure constants: 
$ 
 \left[ {\sigma_A,\sigma_B} \right] = 
    i f_{ABC} \sigma_C 
$, 
where $\sigma_A$ are the generators of the Lie algebra of $G$. 
We choose the $\sigma_A$ to be matrices satisfying 
\be  
  Tr\left( {\sigma_A\sigma_B} \right) = \delta_{AB} 
\quad .  
\label{eq2} 
\ee   

It is straightforward to quantize the gauge system 
governed by 
Eq.\bref{eq1}. 
The hamiltonian is 
\be 
  H = {1 \over 2}\pi_A\pi_A + 
   igf_{ABC} \bar\psi_A\phi_B\psi_C 
\quad ,   
\label{eq3} 
\ee   
where $\phi_A$ and $\pi_B$, 
as well as $\psi_A$ and $\bar\psi_B$ 
are conjugate variables 
satisfying 
$  
  \left[ {\phi_A,\pi_B} \right] = 
    i\delta_{AB} 
$ 
and 
$ 
  \left\{ {\psi_A, \bar\psi_B} \right\} = 
   \delta_{AB} 
$. 
States $\state$ 
must satisfy the Gauss law constraints  
\be  
 G_A \state = 0 
\quad ,   
\label{eq4} 
\ee   
where  
\be  
  G_A = 
  f_{ABC}\left( {\phi_B \pi_C - 
           i\bar\psi_B \psi_C} \right) 
\quad .  
\label{eq5} 
\ee   
In other words, 
states must be gauge invariant. 
The degrees of freedom $A_B$ do not enter the hamiltonian --  
their role has been replaced by these Gauss law constraints. 

Note that 
\be  
  H = 
  {1 \over 2}\pi_A\pi_A + g\phi_A G_A 
\quad ,   
\label{eq6} 
\ee    
so that, on gauge-invariant states, 
$H$ reduces to 
${1 \over 2}\pi_A\pi_A$. 
If it were not for 
Eq.\bref{eq4}, 
the theory would be free. 
One needs only to find the gauge-invariant 
plane wave states $\state$: 
\be  
    {1 \over 2} \pi_A\pi_A \state = E_s \state 
\ , \quad \quad  G_A \state = 0  
\quad .    
\label{eq7} 
\ee     

The lagrangian 
in Eq.\bref{eq1} 
is invariant under the supersymmetry 
transformations generated by  
$ 
  Q = \psi_A\pi_A 
$ 
and 
$  
 \bar Q = \bar\psi_A \pi_A 
$. 
As usual, 
the anticommutator of $Q$ and $\bar Q$ 
yields the hamiltonian 
up to gauge transformations: 
$ 
  \left\{ {Q, \bar Q} \right\} = 
  \pi_A \pi_A = 2H - 2g \phi_A G_A 
$.  

States are classified 
according to their fermion number. 
In particular, 
one can define the fermion vacuum $\plusstate$ 
to be annihilated by all the $\psi_A$: 
\be  
  \psi_A \plusstate = 0 
\quad . 
\label{eq8} 
\ee     
Assign fermion number $0$ to $\plusstate$. 
All  other fermionic sectors 
are obtained by repeatedly applying $\bar\psi_A$. 
Since there are $n_G$ such fermions, 
there are states with fermion number 
$0$, $1$, $\dots$, $n_G$, 
and the total number of fermionic Fock-space states 
is $2^{n_G}$. 

The case of $G=SU(2)$, 
for which $n_G = 3$,  
has been solved by 
M. Claudson and M. Halpern 
\ct{ch85}. 
They found 
$$ 
  \kstate_0 = {{\sin \left( {kr} \right)} \over {kr}} \plusstate  
\ , 
$$ 
$$ 
  \kstate_1 = {Q \over k}\kstate_0 = 
  \left[ {{{\sin \left( {kr} \right)} \over 
                {\left( {kr} \right)^2}} - 
   {{\cos \left( {kr} \right)} \over {kr}}} \right] 
   {1 \over r}\phi_A\psi_A \plusstate  
\ , 
$$ 
$$ 
  \kstate_2 = 
  \left[ {{{\sin \left( {kr} \right)} \over 
                {\left( {kr} \right)^2}} - 
   {{\cos \left( {kr} \right)} \over {kr}}} \right] 
   {1 \over {2r}}\varepsilon_{ABC} 
        \phi_A \bar\psi_B \bar\psi_C \plusstate  
\ , 
$$ 
\be  
  \kstate_3 = {Q \over k} \kstate_2 = 
   {{\sin \left( {kr} \right)} \over {kr}} 
   \bar\psi_1 \bar\psi_2 \bar\psi_3 \plusstate  
\quad ,  
\label{eq9} 
\ee  
where the subscript $p$ on $\kstate_p$ indicates 
the fermion number.  
Here, $ r = \sqrt{\phi_A \phi_A} $,   
$k$ is any non-negative real number, 
and $\varepsilon_{ABC}$ 
is the completely antisymmetric tensor on three indices. 
The states 
in Eq.\bref{eq9} 
are only plane-wave normalizable, 
as expected, since the spectrum is continuous. 

The goal of this section 
is to obtain zero-fermion-number solutions 
for $G=U(n)$ and $G=SU(n)$. 
We first treat the $U(n)$ case. 
Let $\sigma_A$ be the matrix generators 
in the fundamental representation, 
with a normalization respecting 
Eq.\bref{eq2}. 
Let $\Phi = \phi_A \sigma_A$ be 
an $n \times n$ matrix of scalar fields. 
Since a gauge-invariant functional of the $\phi_A$ 
depends only on the eigenvalues of $\Phi$,  
write 
\be  
  \Phi = \phi_A \sigma_A = U^{-1}DU 
\quad ,  
\label{eq10} 
\ee      
where $D$ is a diagonal matrix of 
the eigenvalues $\lambda_j$ of $\Phi$:   
\be 
 D = \left( \matrix{\lambda_1 \quad \quad 0 \hfill\cr
  \quad \ddots \quad \hfill\cr
  0 \quad \quad \lambda_n\hfill\cr} \right)
\quad ,  
\label{eq11} 
\ee   
and $U$ is some unitary transformation. 
Let us look for solutions of the form 
$f\left( {\lambda_1,\ldots ,\lambda_n} \right) \plusstate$. 
Acting on such a state, 
\be  
  \pi_A\pi_A \to  
  - {1 \over {{\cal M}^2}}\sum\limits_{j=1}^n 
  {{\partial  \over {\partial \lambda_j}}}   
    {\cal M}^2 {\partial  \over {\partial \lambda_j}} 
\quad ,  
\label{eq12} 
\ee  
where ${\cal M}^2$,  
the Vandermonde determinant measure factor, is the square of   
\be  
  {\cal M} = 
    \prod_{1 \le i<j \le n} {\left( {\lambda_i-\lambda_j} \right)} 
\quad .   
\label{eq13} 
\ee  
After some guesswork, 
we have found solutions to 
$ 
  {1 \over 2} \pi_A \pi_A f = E f 
$.  
They are given by 
\be  
  f = 
      {\cal M}^{-1} 
   \exp \left[ {i\sum\limits_{j=1}^n {k_j\lambda_j}} \right] 
\quad ,   
\label{eq14} 
\ee   
where the ``momenta'' $k_j$ are arbitrary real numbers. 
The energy eigenvalue is 
\be 
   E = {1 \over 2} \sum\limits_{j=1}^n {k_j^2} 
\quad . 
\label{eq15} 
\ee  
In verifying 
$ 
  {1 \over 2} \pi_A \pi_A f = E f 
$, 
one needs to used the identity 
\be  
  \sum\limits_{\scriptstyle {k \ne j}\hfill\atop
  {\scriptstyle {k \ne i}\hfill\atop
  \scriptstyle {j \ne i}\hfill}} 
   {{1 \over {\lambda_i-\lambda_j}}
    {1 \over {\lambda_i-\lambda_k}}} = 0 
\quad , 
\label{eq16} 
\ee 
which, incidently, arises 
in obtaining covariant superstring amplitudes 
for fermionic scattering processes 
\ct{kllss87}. 

The solutions 
in Eq.\bref{eq14} 
behave badly when any two eigenvalues approach 
each other. 
It is possible, 
however, 
to obtain regular solutions by taking linear combinations 
of Eq.\bref{eq14}. 
A unique class of regular solutions is achieved  
by antisymmetrization 
using the permutations $\sigma$ 
of the permutation group $S_n$ on $n$ elements: 
\be  
  \kstate_0 = 
  {\cal N} {\cal M}^{-1} 
   \sum\limits_{\sigma \in S_n} 
   {\left( {-1} \right)^\sigma } 
     \exp \left[ {i\sum\limits_j 
    {k_{\sigma \left( j \right)}\lambda_j}} \right] \plusstate  
\quad , 
\label{eq17} 
\ee 
where ${\cal N}$ is a normalization factor 
and ${\left( {-1} \right)^\sigma }$  
is $+1$ for even permutations and $-1$ for odd permutations. 
It is easy to verfiy that $\kstate_0$ 
is non-singular as $\lambda_j \to \lambda_i$. 

When $G=SU(n)$, 
a solution is obtainable  
from the $G=U(n)$ case 
because the system is separable. 
Select the generator index  
for the diagonal $U(1)$ subgroup to be 
the last one, $n^2$.  
For convenience, 
relabel this index as $0$. 
Hence, 
$\sigma_{n^2} = \sigma_0$ 
where 
$  
  \sigma_0 = {I_n / {\sqrt {n}}}  
$ 
and $I_n$ is the $n \times n$ identity matrix.  
Since the laplacian on $U(n)$, as well as the hamiltonian, 
splits into a $U(1)$ part and an $SU(n)$ part as  
\be  
  {1 \over 2}\pi_A\pi_A = 
  {1 \over 2}\pi_0\pi_0 + 
   {1 \over 2}\sum\limits_{A=1}^{n^2-1} {\pi_A\pi_A} = 
    H_{U\left( 1 \right)} + 
    H_{SU\left( n \right)}-\phi_AG_A 
\quad , 
\label{eq18} 
\ee  
solutions factorize into a product of a $U(1)$ wave function 
$f_{U\left( 1 \right)}$ 
times an $SU(n)$ wave function $f_{SU\left( n \right)}$ via    
\be 
  f_{U\left( n \right)}  = 
    f_{U\left( 1 \right)} f_{SU\left( n \right)} 
\quad .  
\label{eq19} 
\ee 
Since 
$f_{U\left( 1 \right)}$ 
is a function of the sum of the eigenvalues 
and  
$f_{SU\left( n \right)}$ 
is a function of differences of eigenvalues,  
write 
\be  
  \lambda_j = 
  \left( {\lambda_j-\Sigma_\lambda } \right) + 
   \Sigma_\lambda  
\ , \quad \quad  
{\rm where} 
\quad 
  \Sigma_\lambda = 
  {1 \over n}\sum\limits_{j=1}^n {\lambda_j} 
\quad .  
\label{eq20} 
\ee  
Performing the factorization 
in Eq.\bref{eq19}, 
one finds 
$$  
  f_{U\left( 1 \right)} = 
   \exp \left[ {in\Sigma_k\Sigma_\lambda } \right] 
\ , 
$$ 
\be   
  f_{SU\left( n \right)} = 
   {\cal N} {\cal M}^{-1} \sum\limits_{\sigma \in S_n} 
    {\left( {-1} \right)^\sigma } 
   \exp \left[ {i\sum\limits_{j=1}^n 
   {k_{\sigma \left( j \right)} 
    \left( {\lambda_j - \Sigma_\lambda } \right)}} \right] 
\quad . 
\label{eq21} 
\ee 
In Eq.\bref{eq21}, 
$k_{\sigma \left( j \right)}$ 
can be replaced by 
$ 
  k_{\sigma \left( j \right)} - \Sigma_k 
$,  
where 
\be 
  \Sigma_k = 
  {1 \over n}\sum\limits_{j=1}^n {k_j} 
\quad , 
\label{eq22} 
\ee 
due to 
$ 
   \sum\limits_{j=1}^n  
     \left( {\lambda_j - \Sigma_\lambda } \right) = 0 
$.   
Hence, 
the $SU(n)$ wave functions really depend on only $n-1$ 
momenta 
since 
$ 
   \sum\limits_{j=1}^n  
     \left( k_{j} - \Sigma_k \right) = 0 
$.   
The energy separates into a $U(1)$-part $E_{U\left( 1 \right)}$ 
and an $SU(n)$-part $E_{SU\left( n \right)}$: 
\be 
   E = 
  {1 \over 2}\sum\limits_{j=1}^n {k_j^2} = 
   {1 \over 2}\bar k_0^2 + 
  {1 \over 2}{{n-1} \over n}\sum\limits_{j=1}^n {\bar k_j^2} = 
      E_{U\left( 1 \right)} + E_{SU\left( n \right)} 
\quad , 
\label{eq23} 
\ee 
where the barred momenta are 
\be 
  \bar k_0 
   \equiv \sqrt n\Sigma_k =  
   {1 \over {\sqrt n}}\sum\limits_{j=1}^n {k_j} 
\ , \quad \quad 
  \bar k_j = 
   \sqrt {{n \over {n-1}}}\left( {k_j-\Sigma_k} \right) 
\quad . 
\label{eq24} 
\ee 
The ${({n-1}) / n}$ in 
$ E_{SU\left( n \right)}$ compensates for 
the one constraint on the $\bar k_j$ of  
$ 
  \sum\limits_{j=1}^n {\bar k_j} = 0 
$. 

\medskip 
{\bf\large\noindent III.\ The Construction of Other Solutions}

It turns out that the $N=2$ system 
has infinitely many operators $O$ that are conserved 
up to Gauss's law. 
These operators can generate new solutions from old ones. 
Let $O$ be gauge invariant
and not a functional of the $\phi_A$. 
Then if $\state$ satifies 
$H \state = E_s \state$, 
for $H$ 
in Eq.\bref{eq6} 
and if $O \state = \stateprime \ne 0$, 
then $\stateprime$ is also a gauge-invariant eigenstate 
of $H$ with the same energy: 
$H\stateprime = E_s \stateprime$. 
The proof is straightforward. 
Constructing $O$ satisfying these criteria 
is simple. 
It suffices to take $O$ to be a trace 
or products of traces 
of the 
$\Pi \equiv \sigma^A \pi_A$ 
and the 
$\bar\Psi \equiv \sigma^A \bar\psi_A$. 
Examples of such operators are 
$$ 
  O_1 = Q = 
    Tr\left( {\bar\Psi \Pi } \right) 
\ , \quad \quad 
  O_2 = 
    Tr\left( {\bar\Psi \bar\Psi \Pi } \right) 
\ , 
$$ 
\be  
  O_3 = 
   Tr\left( {\bar\Psi \bar\Psi \bar\Psi } \right) = 
   {1 \over 2}f_{ABC}\bar\psi_A \bar\psi_B \bar\psi_C 
\quad , 
\label{eq25} 
\ee 
$ 
  O' = 
   Tr\left( {\bar\Psi \Pi \Pi} \right) 
   Tr\left( {\bar\Psi \bar\Psi \Pi } \right) 
$, etc.. 
When these operators act 
on the states $\kstate_0$ 
of Eq.\bref{eq17}, 
energy eigenstates are produced, 
although they might not be new states. 
For example, 
$ 
  Tr\left( {\Pi \Pi } \right) 
$ 
produces the same eigenstate 
up to a factor of $2 E_s$. 
Whether a new eigenstate arises 
also depends on the group $G$: 
When $G=SU(2)$, 
$ 
  Tr\left( {\Pi \Pi \Pi } \right) 
$ 
gives zero because the symmetric 
``$d$ symbol'' vanishes; 
for $ G=SU(n)$ with $n \ge 3$, 
$ 
  Tr\left( {\Pi \Pi \Pi } \right) 
$ 
yields a new state. 
Like the $\kstate_0$, 
the $O$-generated states 
are not normalizable 
because the spectrum is continuous.%
{\footnote{ 
Certain states might be badly non-normalizable. 
The issue of which states should be retained 
in the Hilbert space goes beyond the scope 
of the present work.}}  
Although an infinite number of $O$ can be constructed,  
only a finite number generate independent states. 
It may be that all eigenstates of $H$ 
can be obtained by applying the $O$ 
to the $\kstate_0$. 

Let us verify this conjecture for $G=SU(2)$. 
In doing so, 
we shall also illustrate the reduction procedure 
of Sect.II for going from $U(n)$ to $SU(n)$. 
Factorizing the wave function 
in Eq.\bref{eq17}  
for the $n=2$ case yields 
$$ 
 {1 \over {\lambda_1 - \lambda_2}} 
  \left( {\exp \left[ {ik_1\lambda_1 + ik_2\lambda_2} \right] - 
  \exp \left[ {ik_2\lambda_1 + ik_1\lambda_2} \right]} \right) = 
$$ 
$$ 
  \exp \left[ {{i \over 2}\left( {k_1 + k_2} \right) 
  \left( {\lambda_1 + \lambda_2} \right)} \right] \times  
$$ 
$$ 
  {1 \over {\lambda_1 - \lambda_2}} 
   \left( {\exp \left[ {{i \over 2} 
     \left( {k_1 - k_2} \right) 
   \left( {\lambda_1 - \lambda_2} \right)} \right] - 
     \exp \left[ - {{i \over 2} 
    \left( {k_1 - k_2} \right) 
   \left( {\lambda_1 - \lambda_2} \right)} \right]} \right) 
$$ 
\be 
  = i\sqrt 2 \bar k \exp 
   \left[ {{i \over 2}\left( {k_1 + k_2} \right) 
   \left( {\lambda_1 + \lambda_2} \right)} \right] 
   {1 \over {\bar kr}}\sin \left( {\bar kr} \right) 
\quad , 
\label{eq26} 
\ee 
where 
\be 
  \bar k \equiv {{k_1-k_2} \over {\sqrt 2}} 
\quad , 
\label{eq27} 
\ee  
and 
\be 
  r \equiv \sqrt {\phi_A\phi_A} \Rightarrow 
  r = 
   {{\left| {\lambda_1-\lambda_2} \right|} \over {\sqrt 2}} 
\quad , 
\label{eq28} 
\ee  
which follows from  
$ 
 \Phi = 
   \sigma_A\phi_A 
$ 
after diagonalization: 
$ 
  \Phi \to \phi_3 = 
   { \left( {\lambda_1-\lambda_2} \right) / {\sqrt 2}} 
$ 
with 
$ \phi_1 = \phi_2 = 0 $. 
Letting 
$ 
  {\cal N}^{-1} = {{i\sqrt 2\bar k}} 
$,  
one obtains the wave function in $\kstate_0$ 
in Eq.\bref{eq9} 
since the factorized form 
in Eq.\bref{eq26} 
leads to  
\be 
  f_{U(1)} = 
   \exp \left[ {i\bar k_0 \bar\lambda_0} \right] 
  \ , \quad \quad 
  f_{SU(2)} = 
  {1 \over {\bar kr}}\sin \left( {\bar kr} \right) 
\quad , 
\label{eq29} 
\ee 
Here, 
$ 
  \bar k_0 = \left({k_1 + k_2}\right) / {\sqrt 2} 
$ 
and 
$  
  \bar\lambda_0 = 
   \left( {\lambda_1 + \lambda_2} \right) /  {\sqrt 2} 
$. 
The energy separates as 
in Eq.\bref{eq23} 
with 
\be  
  E_{U(1)} = 
   {1 \over 2}\bar k_0^2 
\ , \quad \quad  
  E_{SU(2)} = 
   {1 \over 2}\bar k^2 
\quad . 
\label{eq30} 
\ee 
Finally, 
a short calculation shows that when the $O_i$ 
in Eq.\bref{eq25} 
are applied to 
$\kstate_0 = f_{SU(2)} \plusstate$, 
the states $\kstate_i$ 
in Eq.\bref{eq9} 
are generated 
up to an overall normalization. 

\medskip
{\bf\large\noindent IV.\ Perturbative Solutions for $N>2$}

When more than two supersymmetries are present, 
the matrix models no longer appear to be exactly solvable. 
For $N>2$, 
the degrees of freedom are 
a real gauge potential $A_B$, 
a set of real scalar $\phi_B^m$, 
and a set of fermions 
$\psi_B^\alpha$, 
where $m$ and $\alpha$ 
label the different sets. 
Quantization leads to hamiltonians of the 
form\ct{ch85}  
\be 
  H = 
    {1 \over 2} \pi_A^m\pi_A^m + 
    {1 \over 4} g^2\left( {f_{ABC}\phi_B^l\phi_C^m} \right) 
     \left( {f_{ADE}\phi_D^l\phi_E^m} \right) + 
    igf_{ABC} \bar\psi_A^\alpha \phi_B^m 
    \Gamma_{\alpha \beta }^m\psi_C^\beta  
\quad , 
\label{eq31} 
\ee 
the Gauss constraints 
\be 
  G_A = 
     f_{ABC}\left( {\phi_B^m \pi_C^m - 
     i \bar\psi_B^\alpha \psi_C^\alpha } \right) 
\quad , 
\label{eq32} 
\ee 
and the commutation relations 
$  
 \left[ {\phi_A^l,\pi_B^m} \right] = 
   i\delta^{lm} \delta_{AB} 
$ 
and 
$  
  \left\{ {\psi_A^\alpha ,\bar\psi_B^\beta } \right\} = 
    \delta^{\alpha \beta } \delta_{AB} 
$. 
In Eq.\bref{eq31}, 
the $\Gamma_{\alpha \beta }^m$, 
for $m=1, 2, \dots, p$, 
are matrix representations of an $SO(p)$ Clifford algebra. 
For example, 
when $N=4$, 
$p=3$, 
$\alpha = 1$ or $2$, 
and the $\Gamma_{\alpha \beta }^m$  
are $2 \times 2$ Pauli matrices. 
When $N=16$, 
$p=9$, 
$\alpha = 1,2, \dots , 16$,  
the $\Gamma_{\alpha \beta }^m$  
are $16 \times 16$ real matrices 
satisfying 
$\left[ { \Gamma^l , \Gamma^m } \right] = 2 \delta^{lm}$,  
and 
the fermions are real. 
It is possible to organize the $16$ real fermions 
into $8$ complex ones 
at the cost of making less manifest group properties. 

Because the potential-energy terms 
in Eq.\bref{eq31} 
are no longer proportional to Gauss law constraints, 
the full effect of the interactions is felt 
so that the equation  $H \state = E_s \state$ 
is difficult to solve.  
A perturbative approach is possible. 
To begin a perturbative expansion, 
solutions to the $g=0$ system must be known. 
Such solutions are obtainable 
using the methods of Sections II and III 
because, 
when $g$ is zero, 
the hamiltonian is a sum of 
$p$ independent $N=2$ hamiltonians. 
Hence, 
the system factorizes. 
The analog of the state 
in Eq.\bref{eq17} 
is 
\be 
  \kstate_0 = 
   {\cal N} \prod_m  
   \left[ {{\cal M}_m^{-1} \sum\limits_{\sigma \in S_n} 
     {\left( {-1} \right)^\sigma } 
     \exp \left( {i\sum\limits_j {k_{\sigma 
    \left( j \right)}^{(m)}\lambda_j^{(m)}}} \right) 
         } \right] \plusstate 
\quad , 
\label{eq33} 
\ee 
where 
$\plusstate$ is annihilated by all the $\psi_A^\alpha$, 
$k_j^{(m)}$ are momenta for the $m$th sector, 
and   
\be 
  {\cal M}_m^{-1}  \equiv 
  \prod_{i<j} {\left( {\lambda_i^{(m)} - \lambda_j^{(m)}} \right)}^{-1} 
\quad , 
\label{eq34} 
\ee 
where the $\lambda_j^{(m)}$ 
are the eigenvalues of 
the matrix $\phi^m = \sigma^A \phi_A^m$. 
The energy is 
\be 
  E = {1 \over 2} \sum\limits_{j,m} \left( {k_j^{{(m)}}} \right)^2
\quad . 
\label{eq35} 
\ee 
Additional eigenstates 
are generated by applying to $\kstate_0$ 
operators $O$  
that are gauge invariant and that are functionals 
only of the $\pi_B^m$ and fermions. 
Such $O$ involve a trace or products of traces of the 
$\Pi^m \equiv \sigma^A \pi_A^m$ 
and the 
$\bar\Psi^\alpha \equiv \sigma^A \bar\psi_A^\alpha$. 
Examples of such operators are 
$  
    Tr\left( {\Pi^l \Pi^m } \right) 
$,  
$   
    Tr\left( {\bar\Psi^\alpha \bar\Psi^\beta } \right) 
$ for $\alpha \ne \beta$, 
$ 
    Tr\left( {\bar\Psi^\alpha \Pi^m } \right) 
    Tr\left( {\bar\Psi^\alpha \bar\Psi^\beta \Pi^l} \right) 
$,  
etc.. 

\medskip 
{\bf\large\noindent Acknowledgments}  

I thank Bunji Sakita for discussions. 
This work was supported in part 
by the PSC Board of Higher Education at CUNY and 
by the National Science Foundation under the grant  
(PHY-9420615).  

\medskip 

\vfil 

\end{document}